# On a recently proposed dating of the foundation of Julia Augusta Taurinorum (Torino)

**Amelia Carolina Sparavigna**
Politecnico di Torino

Recently, a preprint in arXiv [1] and a series of news announced that Torino, as the roman Augusta Taurinorum, was founded on 30 January 9 BC. The assumption is based on the coincidence of the Julian date, obtained by means of astronomical calculations, and the date of the inauguration, by Augustus, of the Ara Pacis in Rome (historical date). Here we show that, in the framework proposed in [1], the abovementioned date of the foundation is impossible. The impossibility is based on the following fact. The Julian date (30 January 9 BC), proposed by the authors in arXiv [1], cannot definitely be the historical 30 January 9 BC of the Julian Calendar.

Here we comment the results proposed in a preprint published in arXiv [1]. The preprint is entitled "Dating the foundation of Augusta Taurinorum ex sole. The augustean propaganda and the role of Astronomy". In this preprint, written by S. Caranzano and M. Crosta, the authors tell that their joint study of astronomy and archaeology allowed them to define the foundation date of the city of Turin as the Roman colony Iulia Augusta Taurinorum.

The abstract stresses that the multidisciplinary research, proposed in [1], represents "a new reading of the historical-archaeological sources and the use of astronomy according to the Etruscan-Latin gromatica". Actually, the authors are referring to a ritual of Etruscan origin, mentioned by some Gromatici Veteres [2], and reported by some scholars (see for instance [3,4]). During this ritual, the founder of a Roman colony was orienting the centuriation, that is, the local land subdivision, according to the direction of the sunrise. In particular, the main axis of the centuriation, the Decumanus Maximus, was determined to have the same direction of the rising sun. After the determination of the Decumanus, a perpendicular line, the Cardo, and a series of parallel and perpendicular lines were used to subdivide the land. In [1], it is claimed a new reading of this ritual. However, the method of using astronomical data to find the day of the foundation, according to this ritual, had been already proposed in [5,6].

To determine the date of the foundation, the authors in [1] used the "apparent motion of a True Sun, the possible measurement errors, the atmospheric refraction and the elevation of the horizon, and the Julian date in use in astronomy". A numerical program was used to "define the coincidences of the calendars between the azimuth of the main road axis and the course of the sun". According to the used data, the authors claimed the following days as suitable for the foundation: 30 January and the interval 11, 12 and 13 November. In [5,6], I proposed the days about 30 January and 10 November, stressing that it is necessary to consider the natural horizon. Of course, method can be improved.

As discussed in [6], we can suppose the existence of a link between the foundation of a colony and some Roman festivals [7], and try to find the festival close to the possible day of the foundation. In [1], the authors use the same approach. However, they do not consider the days of November and point out 30 January (Julian date) as the date of foundation. They clearly announce in the abstract [1] the following. "For the series of very particular historical and contextual conditions it was therefore possible to trace with sufficient accuracy the day and the year of foundation of the city: January 30, 9 BC, which coincides, not surprisingly, with a particularly important anniversary for Ottaviano Augusto, Emperor from 27 BC to 14 AC" [1]. Press and television used the clear statement in the abstract [1] to announce that the birthday of Torino had been determined as 30 January 9 BC.

The authors of [1] continue telling, in the abstract, that Torino, founded as Augusta Taurinorum, "was inaugurated on the day of the anniversary of the feast of Pax, established by Augustus and celebrated at the Ara Pacis in Rome starting from 9 BC". Actually, the name of the roman colony is coming from epigraphic sources and from literature. Different names of the colony are given: "Julia Augusta Taurinorum", "Augusta Taurinorum", "Taurinis". It is also supposed that it had two foundations, one during republican or triumvirate period, the other under Augustus [8,9]. Let us stress that no Latin literature exists mentioning the foundation of Torino.

Therefore, in [1], Caranzano and Crosta decided that the day of the foundation was January 30, 9 BC, because the date of 30 January (Julian date), obtained from astronomical calculations, is the same of 30 January (historical date), Festival of the Peace, established by Augustus in 9 BC.

Here we have the clear inconsistency of the assumption of coincidence. Let us stress that the Julian date is a manner of measuring the time used in astronomy, which is different from the Julian Calendar. The date of the Festival of the Peace is that of the inauguration of the Ara Pacis in Rome. From Augustus, we know the year, from Ovid and the Fasti Praenestini the day of the festival [10]. Therefore, this day is given by a historical date of the Julian Calendar of the time. This Calendar is the Calendar proposed by Julius Caesar in 46 BC, to substitute the Numa Calendar. It started from January 1, 45 BC (historical date). For many years the calendar operated by adding a leap year on a cycle of three years instead of four years. In 8 BC, Augustus stopped the intercalation. In 8 AD, the Julian Calendar was in agreement to the astronomical time, and started operating on a cycle of four years. Of the Julian Calendar, I discussed in [11] (see references therein). In my opinion, which is also the opinion of other scholars, the Calendar of Julius Caesar started on January 2 (Julian date, 45 BC).

For what concerns the historical date of 30 January, that of the festival of the Peace, year 9 BC, it has to correspond to 2 or 3 of February 9 BC (Julian date). The difference, that is 2 or 3 of February, is because some scholars are supposing the Julian Calendar as starting on the new moon of January 45 BC, others the day before. In any case, the coincidence of the Julian date and historical date mentioned in [1] is impossible. The difference is large.

As a conclusion, the Julian date (30 January 9 BC), proposed by the authors in arXiv [1], cannot definitely be the historical 30 January 9 BC of the Julian Calendar. The coincidence does not exists. The conclusion in [1], that Augusta Taurinorum was founded on 30 January 9 BC, is wrong.

For other observations on [1], see please Ref.12.